\documentstyle[a4,12pt,epsfig]{article}

\newcommand{\beq}{\begin{equation}}
\newcommand{\eeq}{\end{equation}}
\newcommand{\bea}{\begin{eqnarray}}
\newcommand{\eea}{\end{eqnarray}}

\def\gev{\;{\rm GeV}}

\def\cM{{\cal M}}
\def\cO{{\cal O}}

\def\matz   {|\overline{\cal{M}}_3|^2}
\def\wh{\widehat}
\def\wt{\widetilde}

\def\lapprox{\lower .7ex\hbox{$\;\stackrel{\textstyle <}{\sim}\;$}}
\def\gs     {g_s}

\def\shift  {\rule[-3mm]{0mm}{8mm}}
\def\dps    {\displaystyle}

\def\gapprox{\lower .7ex\hbox{$\;\stackrel{\textstyle >}{\sim}\;$}}

\begin{document}
\titlepage
\begin{flushright}
{DTP/97/102}\\
{hep-ph/9712314}\\
{December 1997}\\
\end{flushright}

\begin{center}

\vspace*{2cm}

{\Large {\bf Radiation Zeros in High-Energy \\[2mm]
$e^+e^-$ Annihilation into Hadrons}}\\

\vspace*{1.5cm}
M.~Heyssler$^{a}$ and W.J.~Stirling$^{a,b}$ \\

\vspace*{0.5cm}
$^a \; $ {\it Department of Physics, University of Durham,
Durham, DH1 3LE }\\

$^b \; $ {\it Department of Mathematical Sciences, 
University of Durham, Durham, DH1 3LE }

\end{center}

\vspace*{4cm}

\begin{abstract}
The process  $e^+e^- \to q \bar q \gamma$  contains  radiation  zeros,
i.e.  configurations  of the  four--momenta  for which the  scattering
amplitude  vanishes.  We calculate  the  positions  of these zeros for
$u$--quark and  $d$--quark  production and assess the  feasibility  of
identifying  the zeros in experiments at high energies.  The radiation
zeros are shown to occur also for massive  quarks, and we discuss  how
the $b \bar b  \gamma$  final  state may  offer a  particularly  clean
environment in which to observe them.
\end{abstract}

\newpage


\section{Introduction}
\label{sec:intro}

In  certain   high--energy   scattering  processes  involving  charged
particles  and the  emission of one or more  photons,  the  scattering
amplitude  vanishes for particular  configurations of the final--state
particles.  Such  configurations are known as {\it radiation zeros} or
{\it null zones}.  A very clear and comprehensive  review by Brown can
be found in Ref.~\cite{Bro95}.

Radiation  zeros have an  interesting  history.  Although  they are in
principle present in QED amplitudes, they first attracted  significant
attention  in  processes  involving  weak  bosons.  For  example,  the
pioneering  papers of Mikaelian,  Sahdev and Samuel  \cite{Mik79}  and
Brown, Sahdev and Mikaelian \cite{Mik79}  considered radiative charged
weak  boson  production  in $q \bar q $ and  $\nu e$  collisions.  The
cross sections for these  processes  vanish when the photon is emitted
in certain  directions (see below).  Recently,  experimental  evidence
for zeros of this type has been found at the Fermilab Tevatron $p \bar
p$  collider   \cite{CDF97}.  In  addition  to  the   phenomenological
analyses, a deeper  theoretical  understanding  was  developed  in the
papers  of  Ref.~\cite{Bro82}:  the  vanishing  of  the  (tree--level)
scattering  amplitude  can be  understood  as  arising  from  complete
destructive  interference of the classical  radiation  patterns of the
incoming and outgoing charged particles.

There have been many  other  studies  exploring  the  phenomenological
aspects  of  radiation   zeros.  For  example,  the   introduction  of
(non--gauge)  `anomalous  couplings'  destroys the cancellation  which
leads to the vanishing of the amplitude, and so radiation zeros can be
used      as      sensitive       probes      of      new      physics
[\ref{ref:Bau94}--\ref{ref:Abr97}].  More recently, the possibility of
detecting  radiation  zeros  in  $eq$  scattering  at  HERA  has  been
investigated [\ref{ref:Bil85}--\ref{ref:Hey97b}].

In the course of analysing the $eq \to eq\gamma$  matrix  elements for
`standard' radiation zeros in  Ref.~\cite{Hey97b},  a new type of zero
was  discovered.  This appears to arise in a wider class of processes,
and in particular in the crossed process $e^+ e^- \to q\bar q \gamma$.
This  opens up the  possibility  of  identifying  radiation  zeros  in
high--energy  $e^+e^-$  annihilation  into  hadrons,  for example at a
future  linear  collider.  The  purpose  of the  present  study  is to
calculate the position of these zeros and to assess the feasibility of
their observation in experiment.

Before studying the $e^+e^-$ annihilation process in detail, it may be
useful to make  some  general  observations  on the  various  types of
radiation  zeros.  The  discussion  is  particularly  simple  when one
considers the  amplitude for the emission of a single soft photon in a
scattering  process  $  1+2  \to  3 + 4 +  \ldots$  involving  charged
particles.

The matrix element for one (soft) photon emission can be written as
\beq
\cM_\gamma\; \simeq \; e  J\cdot \epsilon \; \cM_0 \; ,
\label{eq:me}
\eeq
where  $\cM_0$  is the  leading--order  (no  photon  emission)  matrix
element and $\epsilon$ is the polarisation  vector of the photon, with
polarisation  and  helicity  labels  suppressed  for the  moment.  The
current is given by
\beq
J^\mu = \sum_i e_i\eta_i  { p_i^\mu \over p_i \cdot k} \; , 
\label{eq:current}
\eeq
where $e_i$ is the charge of the $i$th particle and $\eta_i$ = $+1,-1$
for  incoming,   outgoing   particles.  Energy--momentum   and  charge
conservation  give $\sum_i \eta_i p_i^\mu = 0$ and $ \sum_i e_i \eta_i
= 0$ respectively.

The classical (type~1) radiation zeros are obtained by noting that the
condition
\beq
{ e_i \over  p_i \cdot k} = \kappa \; ,
\label{eq:kappa}
\eeq
where $\kappa$ is a constant  independent of $i$,  immediately  yields
$J^\mu  = 0$ and  hence $  \cM_\gamma=  0$,  for  all  helicities  and
polarisations.  Note that type 1 zeros  require all  particles to have
the same sign of electric  charge.  A simple  example is  provided  by
$u(p_1) + \bar d (p_2) \to  W^+(p_3)  + \gamma  (k) $, where a zero of
the amplitude is obtained for
\beq
\frac{2}{3}\; \frac{1}{p_1 \cdot k} = \frac{1}{3}\; 
\frac{1}{p_2 \cdot k} 
\quad \Rightarrow \quad \cos\theta_\gamma = -\frac{1}{3} \; ,
\label{eq:udW}
\eeq
where  $\theta_\gamma$  is the polar angle of the photon in the c.m.s.
frame  with  $\theta_\gamma  =  0^\circ$  in the  incoming  $u$--quark
direction.

Type 2 zeros, on the other  hand, only arise  when the  scattering  is
{\it  planar}  \cite{Hey97b},  i.e.  the  three--momenta  of  all  the
particles  including  the photon lie in the same plane.  In this case,
if one chooses one of the photon polarisation vectors $\epsilon_\perp$
to be orthogonal to the scattering  plane then  $\epsilon_\perp  \cdot
p_i = 0$ for all $i$ gives  $\epsilon_\perp\cdot  J = 0$ for {\it any}
orientation of the particles and photon in the plane.  The requirement
that the amplitude vanishes for {\it all} helicities and polarisations
means that one must also have  $\epsilon_\parallel\cdot  J = 0$, where
(the spatial part of)  $\epsilon_\parallel^\mu$  is in the  scattering
plane  and  orthogonal  to  the  photon  direction.  The  solution  of
$\epsilon_\parallel\cdot  J = 0$ then  gives  the  position  in photon
angular  phase of the radiation  zero.  If we denote the  direction of
the  three--momentum  of particle $i$ by $\vec{n}_i$ and the direction
of the  photon  by  $\vec{n}$,  then the  condition  is (for  massless
particles)
\beq
\sum_i e_i\eta_i  { \vec{\epsilon}_\parallel\cdot\vec{n}_i 
\over   1 -  \vec{n}\cdot\vec{n}_i} = 0 \; , 
\label{eq:t2zero}
\eeq
with  $\vec{\epsilon}_\parallel\cdot\vec{n} = 0$.  After some algebra,
Eq.~(\ref{eq:t2zero}) can be cast into the simpler form
\beq
\sum_i e_i\eta_i  \cot(\theta_{\gamma i}/2) = 0 \; , 
\label{eq:t2zerobis}
\eeq
where $\theta_{\gamma i}$ is the angle between the photon and particle
$i$ directions.\footnote{Note that $\theta_{\gamma i}$ must be defined
in the same  sense  (clockwise  or  anticlockwise  from  the  $\gamma$
direction)  for each  particle,  so that the  $\cot$  can have  either
sign.}  Eq.~(\ref{eq:t2zerobis})  allows  us to  derive  an  existence
proof for the zeros.  First we note that $\cot(\theta_{\gamma i}/2)\to
\infty  $ as  $\theta_{\gamma  i}  \to  0$ ---  these  are  the  usual
collinear   singularities  for  massless  gauge  boson  emission  from
massless  fermions.  Second, we note that not all the $e_i\eta_i $ can
have the same sign (charge  conservation).  Therefore  there exists at
least one angular sector, between $j$ and $k$ say, where the collinear
singularity  has the  opposite  sign (i.e.  $\to \pm  \infty$)  on the
boundaries   of   the   sector.   Since   the   left--hand   side   of
(\ref{eq:t2zerobis}) defines a continuous function of the photon polar
angle  away  from  the  collinear   singularities,  according  to  the
Intermediate  Value Theorem the function must vanish  somewhere in the
sector  between $j$ and $k$.  The exact  location of the zero  depends
not  only  on  the  strength  of  the   collinear   singularities   at
$\theta_{\gamma  j},  \theta_{\gamma  k}=  0$ but  also  on the  other
non--singular  contributions  ($i \neq  j,k$) to the  current  in that
region.

For  $2\to  2$  scattering  the  solutions  to  (\ref{eq:t2zero})   or
(\ref{eq:t2zerobis})  can be found  analytically, for more complicated
scattering  numerical  methods  can be used.  The  existence  of zeros
requires  certain  constraints  on the charges,  masses and scattering
kinematics to be satisfied, as we shall see in the following sections.
For  example,  there  are  no  collinear   singularities  for  massive
fermions,  and  therefore  the  existence  of a radiation  zero in the
angular  sector  depends on how  strongly the  distribution  is peaked
close to the  massive  particles,  which in turn  depends on the exact
value of the mass.

Type 2 zeros do not require  that all the charges  have the same sign.
For  example, the process  $e^- d \to e^-  d\gamma$  has zeros of both
types,  whereas  $e^- e^+ \to d \bar{d}  \gamma$ only has type 2 zeros
(see below).  Although for simplicity we have used soft--photon matrix
elements and kinematics in the discussion  above,  radiation  zeros of
both types are also found when exact  kinematics  and matrix  elements
are used \cite{Hey97b}.

In this paper we present a detailed  theoretical and  phenomenological
study of (type 2) radiation  zeros in the scattering  process $e^- e^+
\to q \bar q \gamma$ at high  energy.  We shall show that zeros  exist
for  both  $u$--  and  $d$--type   quarks  for  all   helicities   and
polarisations.  The  zeros  occur  in  photon   directions  which  are
reasonably  well separated from the directions of the other  particles
in the  scattering.  Unfortunately  it is  very  difficult  to  obtain
analytic  expressions for the positions of the zeros with exact matrix
elements  and phase  space.  Results  for the general  case,  obtained
numerically, will be presented in  Section~\ref{sec:hard}.  However in
the  soft--photon   approximation  (which  in  fact  is  the  dominant
experimental  configuration) it {\it is} possible to obtain reasonably
compact  expressions.  In  Sections~2  and 4 we use  the  soft--photon
approximation  to locate the zeros,  first for  massless  and then for
massive  quarks.  Section~3  briefly  discusses  radiation  at the $Z$
pole.  In Section~\ref{sec:montecarlo} we perform a Monte Carlo study,
based  on the  exact  matrix  elements  and  phase  space,  to  obtain
`realistic'  distributions  of the  type  which  might  be  accessible
experimentally.   Finally,   our    conclusions   are   presented   in
Section~\ref{sec:conc}.


\section{Massless quarks in the soft limit}
\label{sec:softnomass}

We consider the processes
\bea
e^-(1) \; e^+(2) &\longrightarrow& q(3) \; \bar{q}(4) + \gamma(k)\, ,
\label{eq:qqphoton}
\\
e^-(1) \; e^+(2) &\longrightarrow& q(3) \; \bar{q}(4) + g(k)\, .
\label{eq:qqgluon}
\eea
The gluon emission process  (\ref{eq:qqgluon})  does {\it not} contain
radiation zeros, but is useful for comparison.  To begin with we shall
consider   $s$--channel   $\gamma^*$  exchange  only,  as  this  fully
determines  the  positions of the  radiation  zeros.  The exact matrix
elements for these processes are (for massless quarks and leptons, see
for example Ref.~\cite{Ber81})
\bea
\matz(e^-e^+  \rightarrow q\bar{q} + \gamma)&=&
-3e^6e_q^2\; \frac{ {t}^2 +  {t}'^2 +  {u}^2 +  {u}'^2}
{ {s} {s}'}
\left(v_{12} + e_qv_{34}\right)^2\, ,
\label{eq:photonmat}
\\
\matz(e^-e^+  \rightarrow q\bar{q} + g) &=&
-4 e^4 e_q^2 \gs^2\; \frac{ {t}^2 +  {t}'^2 +  {u}^2 +  {u}'^2}
{ {s} {s}' }\left(v_{34}\right)^2\, ,
\label{eq:gluonmat}
\eea
with the standard  definitions  for the  $2\rightarrow  3$  Mandelstam
variables
\bea 
& & {s} = (p_1+p_2)^2 \, , \quad {t} = (p_1-p_3)^2\, , \quad
{u} = (p_1-p_4)^2\, ,  \nonumber \\
& &  {s}' = (p_3+p_4)^2\, , \quad  {t}' = (p_2-p_4)^2\, , \quad
{u}' = (p_2-p_3)^2\, , 
\eea 
and 
\beq v_{ij} = \frac{p_i^\mu}{p_i\cdot k} -
\frac{p_j^\mu}{p_j\cdot k}\, .  
\label{eq:vdef} 
\eeq

\begin{figure}[t]
\begin{center}
\mbox{\epsfig{figure=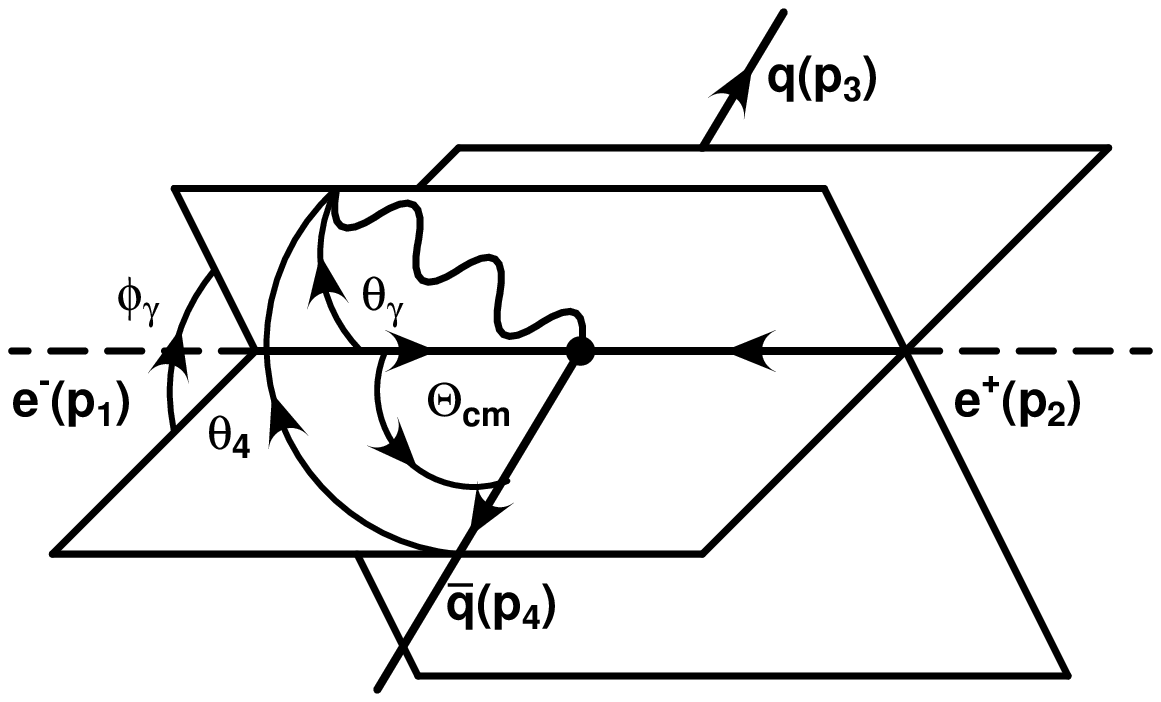,width=11.5cm}}
\caption[]{
Parameterisation of the kinematics for $e^-(p_1) e^+(p_2)  \rightarrow
q(p_3)  \bar{q}(p_4)  + \gamma(k)$  scattering in the $e^-e^+ $ c.m.s.
frame.  The orientation of the photon relative to the scattering plane
is denoted by the angles $\theta_\gamma$ and $\phi_\gamma$.  Note that
$\theta_\gamma = \theta_2$.
}
\label{fig:kinemat}
\end{center}
\end{figure}

In the soft limit, i.e.  $\omega_{\gamma,g}/E_i \rightarrow 0$, we may
use  $2\to  2$  kinematics  for the  $e^-e^+\to  q \bar q$ part of the
process.  The  four--vectors  in the c.m.s.  frame can then be written
as
\bea
p_1^\mu &=& \frac{\sqrt{ {s}}}{2} \left( 1, 0, 0, -1 \right)\, ,
\\
p_2^\mu &=& \frac{\sqrt{ {s}}}{2} \left( 1, 0, 0, 1 \right)\, ,
\\
p_3^\mu &=&  \frac{\sqrt{ {s}}}{2} \left(1, -\sin\Theta_{\rm cm}, 0, 
- \cos\Theta_{\rm cm} \right)\, ,
\\
p_4^\mu &=& \frac{\sqrt{ {s}}}{2} \left(1, \sin\Theta_{\rm cm}, 0, 
\cos\Theta_{\rm cm} \right)\, ,
\\
k^\mu &=& \omega_{\gamma,g} \left(1, \sin\theta_{\gamma,g} 
\cos\phi_{\gamma,g}, 
\sin\theta_{\gamma,g}\sin\phi_{\gamma,g},
\cos\theta_{\gamma,g} \right)\, .
\eea
These kinematics  are illustrated in Fig.~\ref{fig:kinemat}.

Radiation  zeros  for  process   (\ref{eq:qqphoton})  arise  from  the
vanishing of the  $\left(v_{12} +  e_qv_{34}\right)^2$  term.  This is
the  `antenna  pattern'  ${\cal{F}}^{\gamma}  = -J\cdot J$ of the soft
emission           process,          see          for          example
Refs.~[\ref{ref:DKT}--\ref{ref:BOOK}].  A useful  parameterisation  is
to introduce the variables  $z_i =  \cos\theta_{i}$  which specify the
angular separation of the soft photon or gluon from particle $i$.  The
eikonal factors which make up the antenna pattern are then
\beq \label{eq:eikonal}
[ij] \equiv \frac{p_i\cdot p_j}{(p_i \cdot k)(p_j \cdot k)} = 
\frac{1}{\omega^2}\frac{1-\cos\theta_{ij}}{(1-z_i)(1-z_j)}\, ,
\eeq
and the  antenna  patterns  themselves  can be readily  obtained  from
Eqs.~(\ref{eq:photonmat},\ref{eq:gluonmat})
\bea
\frac{1}{2}{\cal{F}}^\gamma &=& [12] + e_q^2[34] -  
e_q\left( [13] + [24] - [14] - [23] \right)\, ,
\label{eq:phoant}
\\
\frac{1}{2}{\cal{F}}^g &=& [34]\, .
\label{eq:gluant}
\eea
We see  immediately  that  there  are no  radiation  zeros  of  type~1
\cite{Hey97b},   as  this  would   require  (for  the   vanishing   of
${\cal{F}}^\gamma$)
\beq
\frac{-1}{1+z_2} = \frac{1}{1-z_2} = \frac{e_q}{1+z_4} = 
\frac{-e_q}{1-z_4} \, ,
\label{eq:contype1}
\eeq
which has no solutions in the physical domain.

In  contrast,  type~2   radiation  zeros  are  located  in  the  event
scattering  plane  \cite{Hey97b}  and do {\em not}  fulfill  condition
(\ref{eq:contype1}).  For a complete set of kinematic variables in the
soft--photon limit we may take the $q \bar q$ c.m.s.  scattering angle
$\Theta_{\rm  cm}$ and two of the $z_i$  variables  introduced  above:
${\cal{F}}^\gamma = {\cal{F}}^\gamma  (\Theta_{\rm  cm},e_q,z_2,z_4)$,
since  $z_1 = -z_2$ and $z_3 = -z_4$ in the  c.m.s.  frame.  To locate
the zeros we solve
\beq
{\cal{F}}^\gamma (\Theta_{\rm cm},e_q,z_2,z_4) = 0
\label{eq:solve}
\eeq
and find
\beq
\hat{z}_4 = -e_q z_2 \pm \sqrt{{ f}(\Theta_{\rm cm},e_q)}\, ,
\label{eq:z4solution}
\eeq
with
\beq
{ f}(\Theta_{\rm cm},e_q) = 1 + e_q^2 + 2e_q\cos\Theta_{\rm cm}\, .
\label{eq:function}
\eeq

As we  expect  the  (type~2)  radiation  zeros  to be  located  in the
scattering  plane,\footnote{Note  that it is  straightforward  to show
that there are no additional zeros with  $\phi_\gamma  \neq 0^{\circ},
180^{\circ}$.}  we set  $\phi_\gamma  =  0^{\circ}$  and  derive as an
additional condition
\bea \nonumber
z_4 = \cos\theta_4 &=& \cos(\theta_\gamma - \Theta_{\rm cm}) =
\sin\theta_\gamma \sin\Theta_{\rm cm} + \cos\theta_\gamma
\cos\Theta_{\rm cm}
\\
&=& \sqrt{1-z_2^2}\sin\Theta_{\rm cm} + z_2\cos\Theta_{\rm cm}\, .
\label{eq:cond2}
\eea

The solutions of Eqs.~(\ref{eq:z4solution}) are tangential hyperplanes
to  Eq.~(\ref{eq:cond2})  in the $\Theta_{\rm cm},z_2$ space for given
charge $e_q$.  Thus we find the positions of the  radiation  zeros for
given $e_q$ and c.m.s.  scattering angle by solving
\beq
\frac{d}{dz_2} \hat{z}_4 = \frac{d}{dz_2}
\left( \sqrt{1-z_2^2}\sin\Theta_{\rm cm} + z_2\cos\Theta_{\rm cm} 
\right)\, ,
\label{eq:differ}
\eeq
which immediately yields
\beq
e_q =  \frac{z_2}{\sqrt{1-z_2^2}} \sin\Theta_{\rm cm} - 
\cos\Theta_{\rm cm}\, .
\label{eq:solver}
\eeq
The solutions are
\beq
\hat{z}_2 = \cos\wh{\theta}_\gamma = \pm \; \frac{e_q + 
\cos\Theta_{\rm cm}}
{\sqrt{{ f}(\Theta_{\rm cm},e_q)}} \, ,
\label{eq:razloc}
\eeq
with  `$+$'  if   $\wh{\phi}_{\gamma}   =  0^{\circ}$   and  `$-$'  if
$\wh{\phi}_{\gamma}   =  180^{\circ}$.  Eq.~(\ref{eq:razloc})   yields
physical  solutions  for  both  $e_q=-1/3$   ($d$--type   quarks)  and
$e_q=2/3$  ($u$--type  quarks) in the complete  range of  $\Theta_{\rm
cm}$.

We  mention several other  interesting  features. 
\begin{itemize}
\item[(i)] 
If we substitute the solution for $\hat{z}_2$ of Eq.~(\ref{eq:razloc})
and $\hat{z}_4$ of Eq.~(\ref{eq:z4solution})  into the antenna pattern
${\cal{F}}^\gamma$ we find
\beq
[12] = e_q^2[34] =  \frac{1}{2}e_q([13]+[24]-[14]-[23])\, ,
\label{eq:cancel}
\eeq
i.e.  the  interference  term  exactly  cancels the sum of the leading
pole terms which are equal.  Therefore solving ${\cal{F}}^\gamma=0$ is
equivalent  to solving  $[12] = e_q^2[34]$  in the massless  case.  We
shall test this feature  later for massive  quarks.  \item[(ii)]  From
Eq.~(\ref{eq:razloc})  we see that the radiation  zeros are orthogonal
to the beam  direction  for  $\cos\Theta_{\rm  cm} = -e_q$ which means
$\Theta_{\rm  cm}  \sim   131.8^{\circ}$   for  $u$--type  quarks  and
$\Theta_{\rm cm} \sim 70.5^{\circ}$ for $d$--type quarks.
\item[(iii)] 
The radiation  zeros are located in different  sectors:  for $d$--type
quarks they are located  between the  directions of the incoming $e^+$
and outgoing $\bar{q}$ and between the incoming $e^-$ and outgoing $q$
directions,  respectively.  For $u$--type  quarks the radiation  zeros
can be found  between the incoming  $e^-$ and outgoing  $\bar{q}$  and
between the incoming $e^+$ and outgoing $q$ directions,  respectively.
This  makes  the  discrimination   between  different  charged  quarks
straightforward, at least in principle.
\item[(iv)]
There is one kinematic  configuration for which the separation between
the radiation zero  direction and the direction of the outgoing  quark
(antiquark) is maximal.  By solving
\beq
\frac{d}{d\Theta_{\rm cm}}\left\{\cos^{-1}
\left( \pm\frac{e_q + \cos\Theta_{\rm cm}}
{\sqrt{{  f}(\Theta_{\rm cm},e_q)}}\right) - 
\Theta_{\rm cm}\right\} = 0\, ,
\label{eq:maxsep}
\eeq
we can show that this is the case if the  radiation  zeros are located
{\em  orthogonal} to the beam direction (the  corresponding  values of
$\Theta_{\rm cm}$ are given above).  The separations are then
\bea
\Delta\theta_\gamma^{\rm max} &=& 41.8^{\circ} 
\qquad\mbox{\rm for $u$--type quarks}\, ,\\
\Delta\theta_\gamma^{\rm max} &=& 19.5^{\circ} 
\qquad\mbox{\rm for $d$--type quarks}\, .
\eea
\end{itemize}
  
\begin{figure}[t]
\begin{center}
\mbox{\epsfig{figure=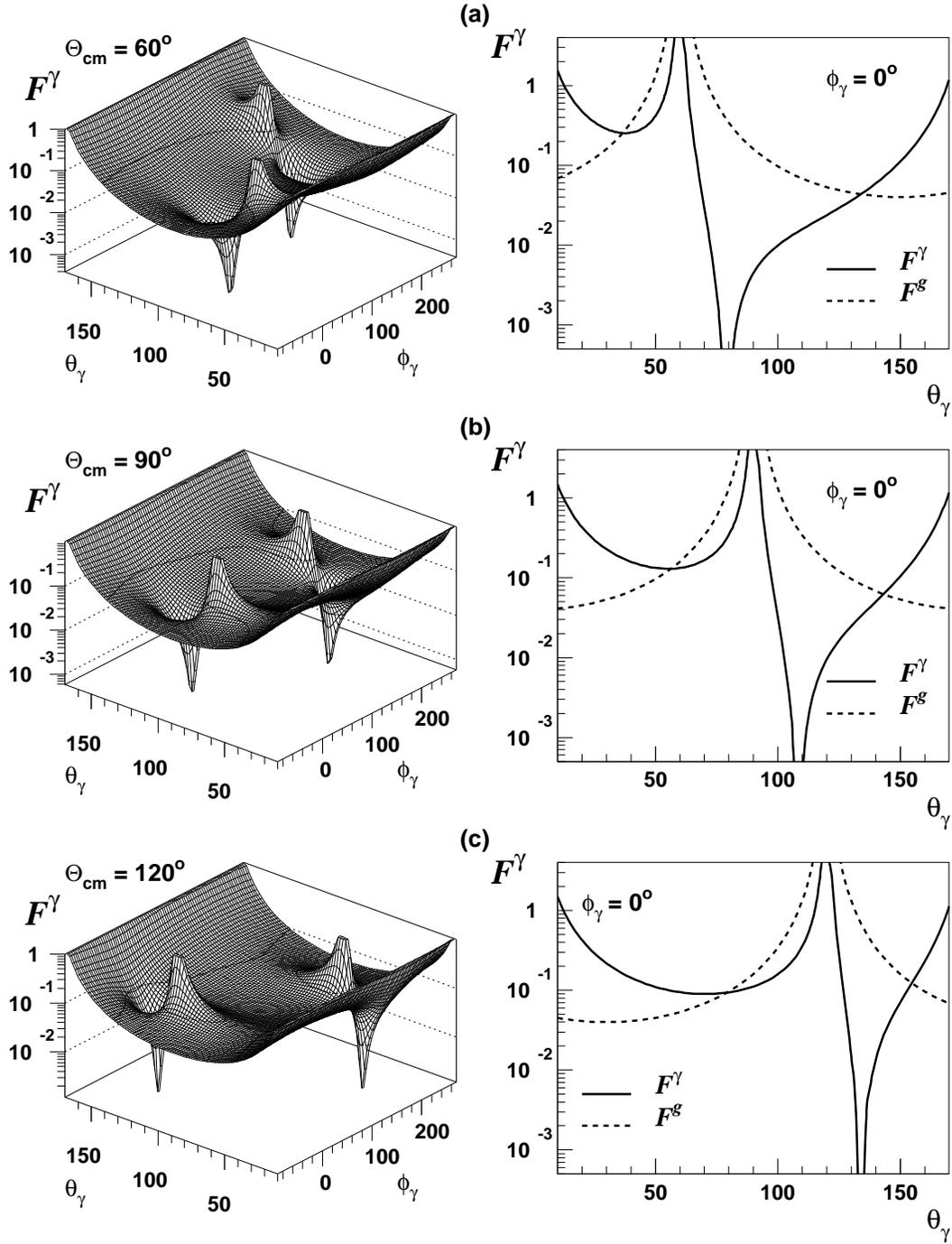,width=16.0cm}}
\caption[]{
Surface plots of the antenna pattern ${\cal{F}}^\gamma$ in the angular
phase space of the soft photon  (left--hand  side) and slices  through
the event plane (right--hand side) at  $\wh{\phi}_\gamma  = 0^{\circ}$
to  illustrate  the  positions of the  radiation  zeros.  We show  the
process $e^-e^+  \rightarrow q_d \bar{q}_d \gamma$ for three different
c.m.s.  frame  angles  (a)   $\Theta_{\rm   cm}  =  60^{\circ}$,   (b)
$\Theta_{\rm   cm}  =   90^{\circ}$   and  (c)   $\Theta_{\rm   cm}  =
120^{\circ}$.  The dashed  lines are the  corresponding  distributions
for soft gluon emission.
}
\label{fig:nomassd}
\end{center}
\end{figure}

\begin{figure}[t]
\begin{center}
\mbox{\epsfig{figure=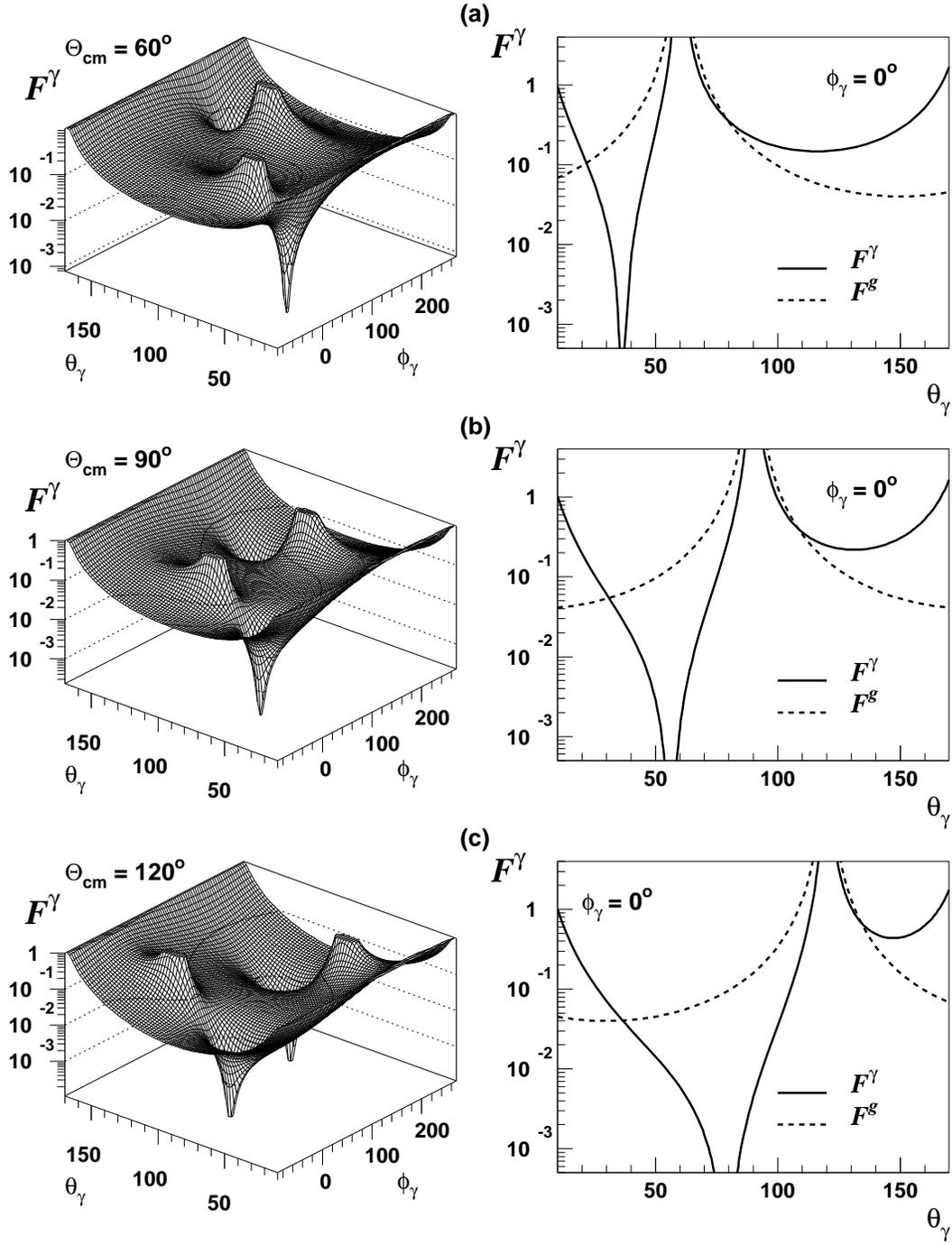,width=16.0cm}}
\caption[]{
Same  as   Fig.~\ref{fig:nomassd},   but  for  the   process   $e^-e^+
\rightarrow q_u \bar{q}_u \gamma$.
}
\label{fig:nomassu}
\end{center}
\end{figure}

In   Figs.~\ref{fig:nomassd},\ref{fig:nomassu}  we  show  the  antenna
patterns   ${\cal{F}}^\gamma$  of  Eq.~(\ref{eq:phoant})  for  process
(\ref{eq:qqphoton})  with  three  different  c.m.s.  frame  scattering
angles  $\Theta_{\rm cm} =  60^{\circ},90^{\circ}$  and $120^{\circ}$.
Additionally we show a slice through the  soft--photon  phase space at
$\wh{\phi}_\gamma=0^{\circ}$   to  illustrate  the  positions  of  the
radiation zeros.  For comparison we also show the antenna patterns for
soft--gluon emission as defined in Eq.~(\ref{eq:gluant}).  This has no
initial--,     final--state     interference    and    therefore    no
zeros.\footnote{Note  that  up  to  charge  factors  the  final--state
collinear   singularities   are  the  same  in  both  cases  however.}
Comparing the  production of $d$--type  quarks and  $u$--type  quarks,
i.e.  Figs.~\ref{fig:nomassd}  and  \ref{fig:nomassu},  shows that the
most striking qualitative feature is the appearance of radiation zeros
in different sectors, as discussed above.

\begin{figure}[t]
\begin{center}
\mbox{\epsfig{figure=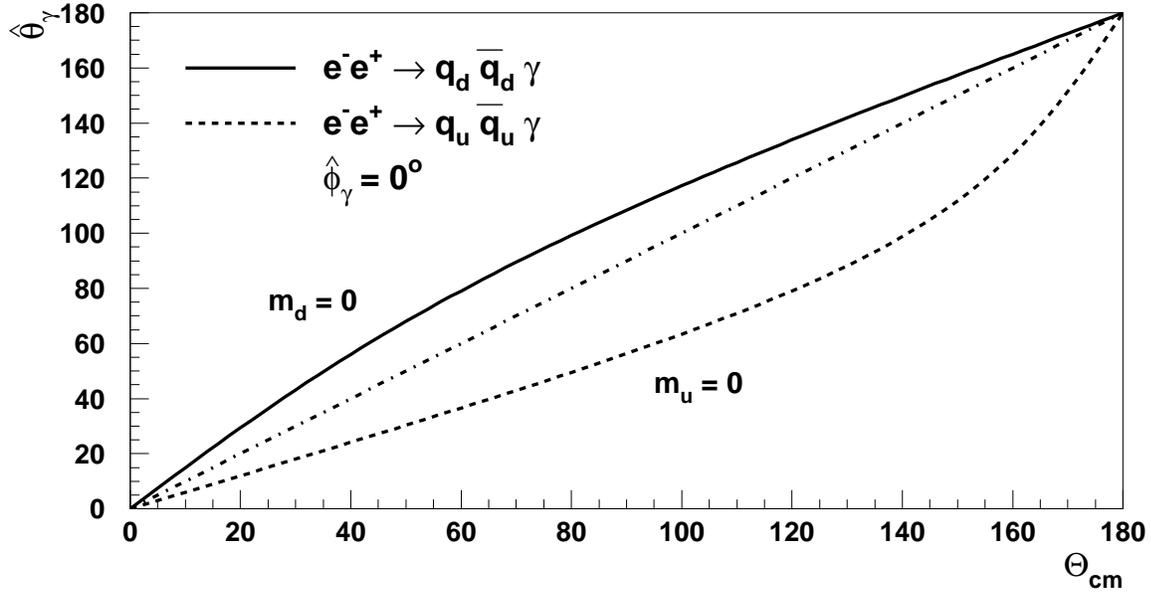,width=16.5cm}}
\caption[]{
The positions $(\wh{\phi}_\gamma,\wh{\theta}_\gamma)$ of the radiation
zeros for the processes $e^-e^+  \rightarrow q_d \bar{q}_d \gamma$ and
$e^-e^+  \rightarrow q_u \bar{q}_u \gamma$ as a function of the c.m.s.
frame scattering angle $\Theta_{\rm cm}$ and fixed $\wh{\phi}_\gamma =
0^{\circ}$.  The   dot--dashed   line  shows  the   position   of  the
final--state   collinear   singularity  (i.e.  the  direction  of  the
outgoing  antiquark).  Massless  quarks  are  assumed.  Note  that the
distribution   for    $\wh{\phi}_\gamma   =   180^{\circ}$   shows   a
$\pi-\Theta_{\rm cm}$ symmetry.
}
\label{fig:nomasspos}
\end{center}
\end{figure}

In  Fig.~\ref{fig:nomasspos} we present the positions of the radiation
zeros   $(\wh{\phi}_\gamma=0^{\circ},\wh{\theta}_\gamma)$   given   by
Eq.~(\ref{eq:razloc}),  as a function of the c.m.s.  frame  scattering
angle, for both  $d$--type and $u$--type  quarks.  Note that radiation
zeros  exist in both  cases for all values of  $\Theta_{\rm  cm}$, and
also  that the  radiation  zeros  for  $u$--type  production  are more
clearly  separated from the collinear  singularities.  For zero--angle
scattering ($\Theta_{\rm  cm}=0^{\circ},180^{\circ}$) the zeros become
pinched along the beam direction.  Note that the $t$--channel  process
$e^+q  \rightarrow  e^+q\gamma$  \cite{Hey97b}  shows a  qualitatively
different behaviour in the zero--angle scattering limit:  in that case
the radiation zeros were located on a cone with fixed angle around the
beam direction.

It  should  be  obvious  from the  above  that in  order to  locate  a
radiation  zero one has to be        able to  distinguish  a quark jet
from an antiquark jet.  Thus if one ($3\leftrightarrow 4$) symmetrises
the expression in  Eq.~(\ref{eq:phoant})  for  ${\cal{F}}^\gamma$, the
interference   term  vanishes  and  there  is  no  zero.  In  practice
distinguishing  between  the quark and  antiquark  jet is likely to be
very  difficult,  but not  impossible.  For example, for  light--quark
jets one could try to tag on the charge of the  fastest  hadron in the
jet.  For heavy (charm,  bottom) quark jets one could in principle use
the charge of the lepton  from the primary  weak decay of the quark to
distinguish  the quark  from the  antiquark.  Methods  like  these are
likely to have poor  efficiency,  so in practice  one would be looking
for a slight dip in the photon  distribution in the vicinity of a zero
when a tagged sample is compared with an untagged sample with the same
overall kinematics.


\section{Radiation on the $Z^0$ pole}

The  general   discussion   on  radiation   zeros   presented  in  the
Introduction  assumed that the hard scattering is  characterised  by a
single  (large)  energy  scale,  so that  the  incoming  and  outgoing
particles  emit photons on the same  timescale.  This  corresponds  to
coherent emission and allows the interference to be maximal.  However,
care must be taken when two  timescales are involved, for example when
there is an intermediate  particle which is relatively long lived.  In
this case the emission off the initial--  and  final--state  particles
can occur at very different  timescales and the  interference  between
them can be  suppressed.  In fact this is exactly what happens for the
process  $e^+e^-  \to f \bar f$ on the $Z$ pole, i.e.  when  $\sqrt{s}
\simeq M_Z$.

A formalism has been  developed for taking these  effects into account
(see   Refs.~[\ref{ref:DKT}   --  \ref{ref:KOhS}]  and  in  particular
Ref.~\cite{KOS}).  In simple terms, the interference  between emission
during the production and decay stages of a heavy  unstable  resonance
of width $\Gamma$ is suppressed by a factor $\chi = \Gamma^2/(\Gamma^2
+ \omega^2)$,  i.e.  there can be no  interference  when the timescale
for  photon  emission  ($\sim  1/\omega$)  is much  shorter  than  the
lifetime of the resonance ($\sim 1/\Gamma$).

In the present  context, the antenna pattern of  Eq.~(\ref{eq:phoant})
is only  valid  far away  from  the $Z$  pole,  $\sqrt{s}  \ll M_Z$ or
$\sqrt{s} \gg M_Z$.  On the $Z$ pole we have, in contrast,
\beq
\frac{1}{2}{\cal{F}}^\gamma_Z = [12] + e_q^2[34] - \chi_Z 
e_q\left( [13] + [24] - [14] - [23] \right)\, ,
\label{eq:Zant}
\eeq
where
\beq
\chi_Z = { M_Z^2 \Gamma_Z^2 \over (P_Z\cdot k)^2 + M_Z^2 \Gamma_Z^2 }
       = { \Gamma_Z^2 \over \omega^2 + \Gamma_Z^2 }\, .
\label{eq:chi}
\eeq
The second  expression  in  (\ref{eq:chi})  corresponds  to the c.m.s.
frame.  For  $\omega  \gg  \Gamma_Z$  there  is no  interference  (and
therefore no radiation zero), and the radiation pattern corresponds to
incoherent emission off the initial-- and final--state  particles.  On
the other hand the  radiation  zero  reappears  in the limit  $\omega/
\Gamma_Z \to 0$.  It is straightforward to show that in this limit the
minimum value of the distribution is $\cO(\omega^2/ \Gamma_Z^2)$.

The  effect  of the  finite  $Z$  width  on the  interference  between
initial--  and  final--state   radiation  was  studied  in  detail  in
Ref.~\cite{JADWAS}.    The    DELPHI    collaboration    \cite{DELPHI}
subsequently confirmed the theoretical  expectations and used the size
of the measured interference to determine $\Gamma_Z$.

Since in the present study we are interested in radiation {\it zeros},
we  must   require   that  the   collision   energy  (and  the  photon
energy\footnote{For  $\sqrt{s} > M_Z$ we can avoid `radiative  return'
to the $Z$ pole by placing an upper bound on the photon  energy.}) are
such that the internal $Z$  propagator is always far off  mass--shell.
This  effectively  guarantees  that  $\chi  = 1$ and  hence  that  the
radiation   pattern   is   again   given   by   Eq.~(\ref{eq:phoant}).
Unfortunately  this  means  that  we are  unable  to use  the  greatly
enhanced  statistics of LEP1 and SLC in searching for radiation zeros.


\section{Massive quarks in the soft photon limit}
\label{sec:massivesoft}

In this section we repeat the analysis of Section~\ref{sec:softnomass}
but now including a non--zero mass for the final--state quarks.  It is
straightforward  to derive the  corresponding  antenna  pattern in the
soft--photon        approximation,        see       for        example
Refs.~[\ref{ref:BOOK},\ref{ref:KOS}--\ref{ref:KOhS}].   The    eikonal
factors for massive particles read
\beq \label{eq:eikonalmass}
[ij]_{m} =  
\frac{1}{\omega^2}\frac{1-\rho_i\rho_j\cos\theta_{ij}}
{(1-z_i\rho_i)(1-z_j\rho_j)}\, .
\eeq
We continue to use massless initial--state  electrons, so that $\rho_1
= \rho_2 =1$ and $\rho_3 = \rho_4 = \rho = \sqrt{  1-4m_q^2/s }$.  The
antenna   pattern   of   Eq.~(\ref{eq:phoant})   now  has   additional
contributions:
\bea \nonumber
\frac{1}{2}{\cal{F}}^\gamma_{m_q} &=& [12]_{m_q} + e_q^2
\left([34]_{m_q} - \frac{1}{2}\frac{m_q^2}{(p_3\cdot k)^2} - 
\frac{1}{2}\frac{m_q^2}{(p_4\cdot k)^2} \right) \\ 
&-& 
e_q\left( [13]_{m_q} + [24]_{m_q} - [14]_{m_q} - [23]_{m_q} \right) .
\label{eq:phoantmass}
\eea
We first consider the limits of $\rho\in[0,1]$
\bea
\rho =0: \qquad {\cal{F}}^\gamma_{m_q} &=& 2[12]_{m_q} = 
\frac{4}{\omega_\gamma^2}\frac{1}{1-\cos^2\theta_\gamma}\, , 
\label{eq:rho0}
\\
\rho =1: \qquad {\cal{F}}^\gamma_{m_q} &=& {\cal{F}}^\gamma\, . 
\label{eq:rho1}
\eea
The first of these  limits is just the  well--known  result that heavy
charged  particles  at rest do not  radiate,  and there are clearly no
radiation zeros.  As ${\cal{F}}^\gamma$  does contain radiation zeros,
we might anticipate a non--trivial $\rho$ dependence of their position
as we increase the mass from zero, with the zeros eventually vanishing
for some critical mass.

A numerical  study  confirms  this result.  We again find zeros in the
scattering        plane        ($\phi_\gamma=0^\circ$).        Solving
${\cal{F}}^\gamma_{m_q}=0$ now gives
\beq
\hat{z}_2^{m_q} = 
\frac{2}{e_q}\frac{e_q\rho\cos\Theta_{\rm cm} + 1 +
\frac{\dps {  g}_\rho(\Theta_{\rm cm},e_q)}{\dps 2 
{f}_\rho(\Theta_{\rm cm},e_q)}}
{\sqrt{-2{  f}_\rho(\Theta_{\rm cm},e_q)
{g}_\rho(\Theta_{\rm cm},e_q)}}\, ,
\label{eq:razlocmass}
\eeq
with
\bea \label{eq:fdef}
{f}_\rho(\Theta_{\rm cm},e_q) &=& 
\rho^2 + e_q^2 + 2e_q\rho\cos\Theta_{\rm cm}\, ,
\\ 
\label{eq:hdef}
{h}_\rho(\Theta_{\rm cm},e_q) &=&
-2e_q\cos\Theta_{\rm cm} \left( 1 + \rho^2 \right) - \rho e_q^2
\left( 1 + \cos^2\Theta_{\rm cm}\right) -2\rho\, ,
\\
\label{eq:gdef}
{g}_\rho(\Theta_{\rm cm},e_q) &=&
\rho {  h}_\rho(\Theta_{\rm cm},e_q) + \rho
\sqrt{{  h}_\rho(\Theta_{\rm cm},e_q)^2 - 
4{f}_\rho(\Theta_{\rm cm},e_q) 
\left( e_q\rho\cos\Theta_{\rm cm} + 
1\right)^2}\, .
\eea
It is  straightforward  to show that in the massless limit  ($\rho=1$)
Eq.~(\ref{eq:razlocmass}) reduces to Eq.~(\ref{eq:razloc}).  Note that
at the positions of the zeros we have, as in the massless case,
\beq
[12]_{m_q} = e_q^2\left([34]_{m_q}-\frac{1}{2}
\left\{ [33]_{m_q}+[44]_{m_q}\right\} \right)\, , 
\label{eq:pole}
\eeq
with the interference again canceling the sum of these two terms.

\begin{figure}[t]
\begin{center}
\mbox{\epsfig{figure=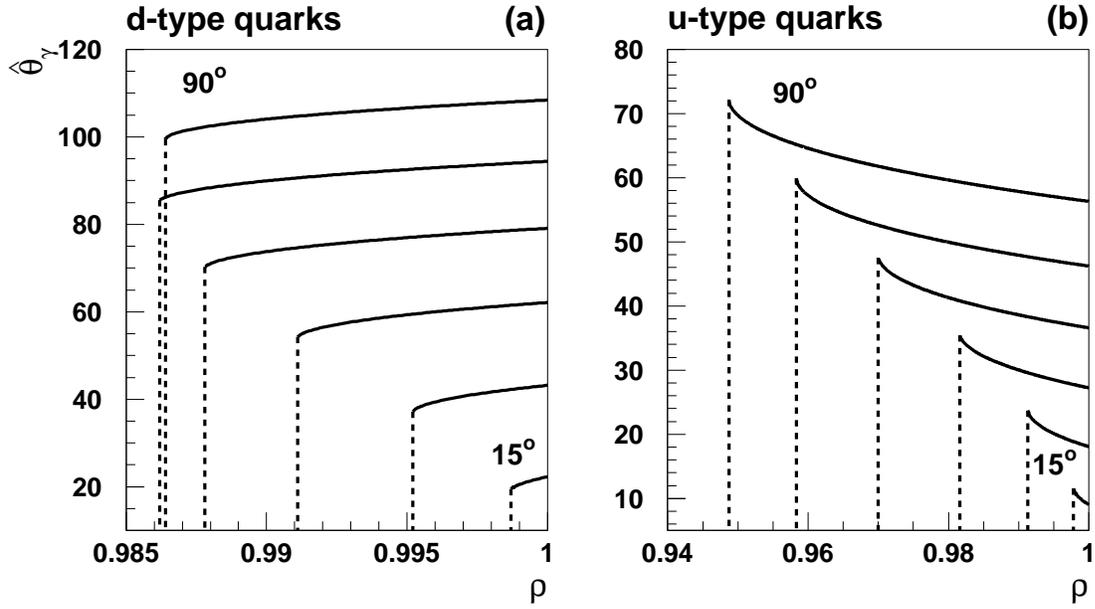,width=16.0cm}}
\caption[]{
The   positions   of  the   radiation   zeros   $(\wh{\phi}_\gamma   =
0^{\circ},\wh{\theta}_\gamma)$  for  massive  quarks  $\rho=\sqrt{1  -
4m_q^2/ {s}}$ (beam energy  $E_{e^-} = 100$~GeV) and different  c.m.s.
scattering     angles     ($\Theta_{\rm     cm}=15^{\circ}-90^{\circ},
\Delta\Theta_{\rm  cm} =  15^{\circ}$).  The  dashed  lines  show  the
values of $\rho_{\rm crit} $.
}
\label{fig:rhocrit}
\end{center}
\end{figure}

\begin{figure}[t]
\begin{center}
\mbox{\epsfig{figure=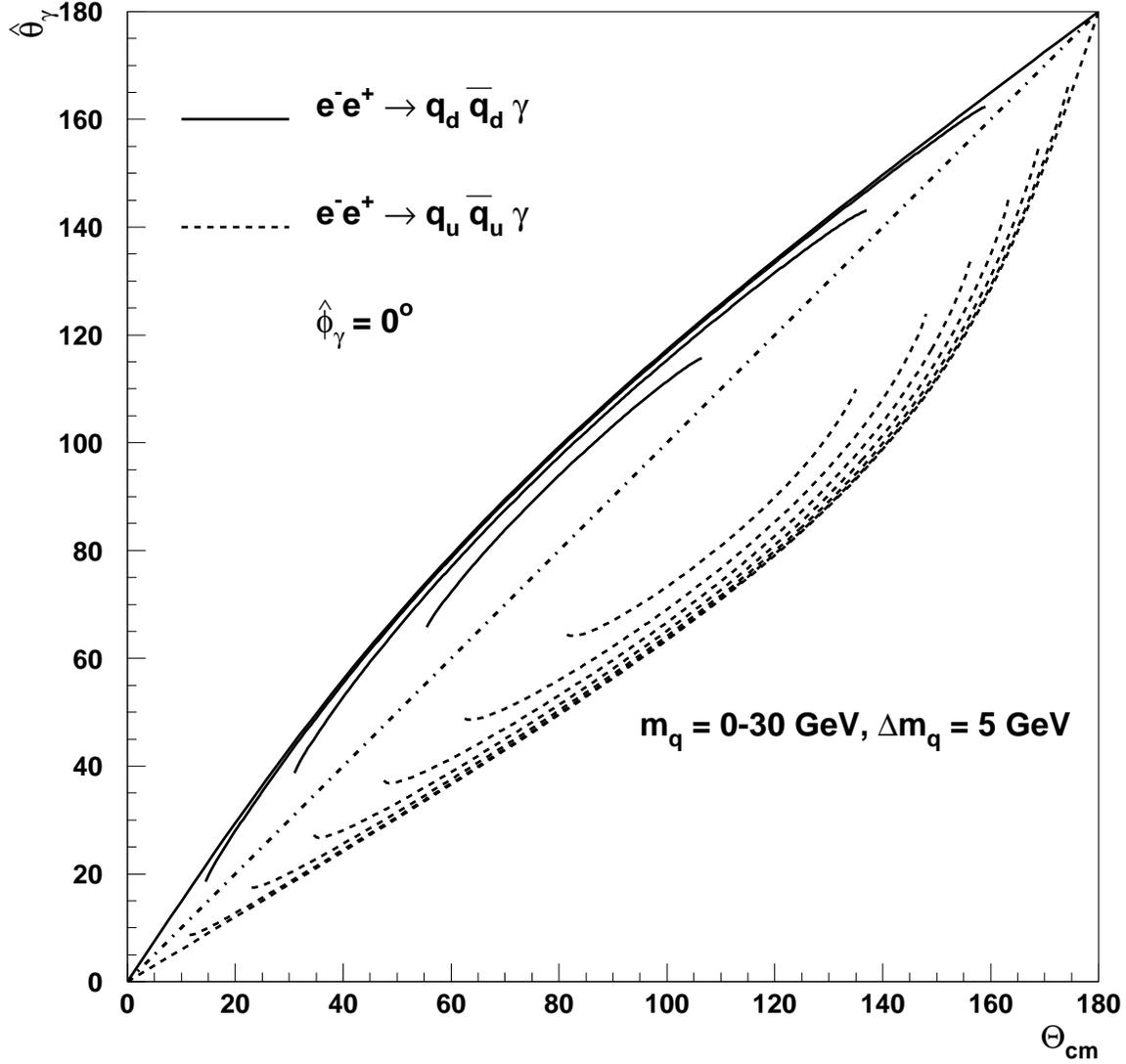,width=16.5cm}}
\caption[]{
Same as Fig.~\ref{fig:nomasspos} but now for massive quarks.  The mass
of the quarks is increased from  $m_q=0$~GeV to $m_q=30$~GeV  in steps
of $\Delta  m_q=5$~GeV.  The higher the mass the closer the  positions
of the zeros move  towards  the  collinear  singularity  (dash--dotted
line).  The beam energy is $E_{e^-} = \sqrt{  {s}}/2 = 100$~GeV.  Note
that the appearance of radiation  zeros is dependent on the quark mass
and the c.m.s.  scattering angle $\Theta_{\rm cm}$.
}
\label{fig:masspos}
\end{center}
\end{figure}

Taken  together,  the  equations  (\ref{eq:razlocmass}--\ref{eq:gdef})
only have physical solutions for a certain range of $\rho\in[\rho_{\rm
crit} ,1]$.  In particular,  if the ratio $  m_q/E_{e^-}$  (quark mass
over beam energy)  becomes too large the  radiation  zeros  disappear.
Fig.~\ref{fig:rhocrit} shows the positions $\wh{\theta}_\gamma$ of the
radiation zeros inside the event plane  ($\wh{\phi}_\gamma=0^{\circ}$)
as a function of $\rho$ for a fixed beam energy $E_{e^-}=100$~GeV, for
both $d$--type and $u$--type  quarks, and for different  values of the
c.m.s.  scattering angle $\Theta_{\rm cm}$.  The dashed lines indicate
the   values   of   $\rho_{\rm   crit}  $.  There  is  one   kinematic
configuration  $\wt{\Theta}_{\rm  cm}$  for  which  $m_q^{\rm  crit} $
becomes  maximal,  i.e.  an upper  limit on the quark  mass for  which
radiation zeros can still be observed.  We find
\beq
\wt{\rho}_{\rm crit}  = \frac{1}{2} \sqrt{4-e_q^2} \quad 
\Longleftrightarrow
\quad \wt{m}_q^{\rm crit}  = \frac{\sqrt{ {s}}}{2}\frac{|e_q|}{2} 
\quad \Longleftrightarrow \quad
\cos\wt{\Theta}_{\rm cm} = \frac{-e_q}{\sqrt{4-e_q^2}}\, .
\label{eq:parmax}
\eeq
For the production of $d$--type quarks we obtain  $\wt{m}_q^{\rm crit}
=16.7$~GeV at $\wt{\Theta}_{\rm cm} = 80.3^{\circ}$, and for $u$--type
quarks we find  $\wt{m}_q^{\rm  crit} =33.3$~GeV at  $\wt{\Theta}_{\rm
cm} = 110.7^{\circ}$.  According to Eq.~(\ref{eq:parmax}) we require a
beam energy of at least $E_{e^-} = 525$~GeV to observe radiation zeros
in the  process  $e^-e^+  \rightarrow  t\bar{t}\gamma$  assuming a top
quark mass of  $m_t=175$~GeV  and an even  higher  energy to achieve a
reasonable    separation    from    the    outgoing    partons    (see
Fig.~\ref{fig:masspos}).\footnote{We   do  not   consider   here   the
contributions  to the radiation  pattern from photon  emission off the
decay products of heavy unstable quarks.}

For $\Theta_{\rm  cm}=90^{\circ}$ we can write the solutions in a very
compact form.  We find as a condition for which radiation zeros exist:
\beq
\rho \geq \rho_{\rm crit}  = \frac{2}{\sqrt{4+e_q^2}}
\quad \Longleftrightarrow \quad
m_q \leq \frac{\sqrt{ {s}}}{2} \frac{|e_q|}{\sqrt{4+e_q^2}}\, .
\label{eq:range}
\eeq
For  example,  in order to  observe  radiation  zeros in  $90^{\circ}$
back--to--back  scattering  with $E_{e^-} = 100$~GeV we need $m_{d{\rm
-type}} <  16.4$~GeV  or  $m_{u{\rm  -type}} <  31.6$~GeV,  conditions
satisfied by all five light--quark flavours.

In  Table~\ref{tab:rhocrit} we present numerical values for $\rho_{\rm
crit}$ and for $m_q^{\rm crit}$, assuming a beam energy for the latter
of   $E_{e^-}=100$~GeV.  The  values   for   $\rho_{\rm   crit}$   are
illustrated in Fig.~\ref{fig:rhocrit}.
\begin{table}[ht]
\begin{center}
\begin{tabular}{|c|c|c|c|c|}\hline
\shift & \multicolumn{2}{c|}{$d$--type quarks} & 
         \multicolumn{2}{c|}{$u$--type quarks} \\ \hline
\shift $\Theta_{\rm cm}$ & $\rho\geq$ & $m_q\leq$ &
                           $\rho\geq$ & $m_q\leq$ \\ \hline\hline
\shift$15^{\circ}$ & 0.9986 &  5.23~GeV  & 0.9977 &  6.72~GeV \\ \hline
\shift$30^{\circ}$ & 0.9951 &  9.82~GeV  & 0.9913 & 13.18~GeV \\ \hline
\shift$45^{\circ}$ & 0.9911 & 13.34~GeV  & 0.9815 & 19.13~GeV \\ \hline
\shift$60^{\circ}$ & 0.9878 & 15.60~GeV  & 0.9699 & 24.34~GeV \\ \hline
\shift$75^{\circ}$ & 0.9861 & 16.59~GeV  & 0.9583 & 28.58~GeV \\ \hline
\shift$90^{\circ}$ & 0.9864 & 16.44~GeV  & 0.9487 & 31.62~GeV \\ \hline
\end{tabular}
\caption[]{
Conditions for the appearance of radiation zeros for different  c.m.s.
frame scattering  angles  $\Theta_{\rm  cm}$.  The numbers in each row
are $\rho_{\rm  crit} $ and $m_q^{\rm  crit} $, assuming a beam energy
of  $E_{e^-}=100$~GeV  for the latter.  Critical mass values for other
beam energies can be obtained by simple rescaling.
}
\label{tab:rhocrit}
\end{center}
\end{table}

An  interesting  conclusion  from   Table~\ref{tab:rhocrit}   concerns
$e^-e^+  \rightarrow  b\bar{b}+\gamma$.  Assuming  a mass  for the $b$
quark  of  $m_b\simeq   4.5$~GeV,  the  actual   kinematics   for  the
observation of radiation  zeros become  critical,  especially at small
c.m.s.  scattering   angles.  For  example,   the   outgoing  $b$  and
$\bar{b}$ jets should be located at around $90^{\circ}\pm  30^{\circ}$
from  the  beam   direction  (cf.  Fig.~\ref{fig:masspos}).  Then  the
radiation zeros not only exist, but are also reasonably well separated
from the collinear  singularities (again cf.  Fig.~\ref{fig:masspos}).


\section{Radiation zeros for arbitrary photon energies}
\label{sec:hard}

We have so far identified radiation zeros using analytic techniques in
the soft--photon  approximation to the scattering  matrix elements and
phase space.  However, as for the $eq \to eq\gamma$ scattering process
studied in Ref.~\cite{Hey97b}, zeros are also found in the {\it exact}
cross  section  for fixed  photon  energies  up to a critical  maximum
value.

\begin{figure}[t]
\begin{center}
\mbox{\epsfig{figure=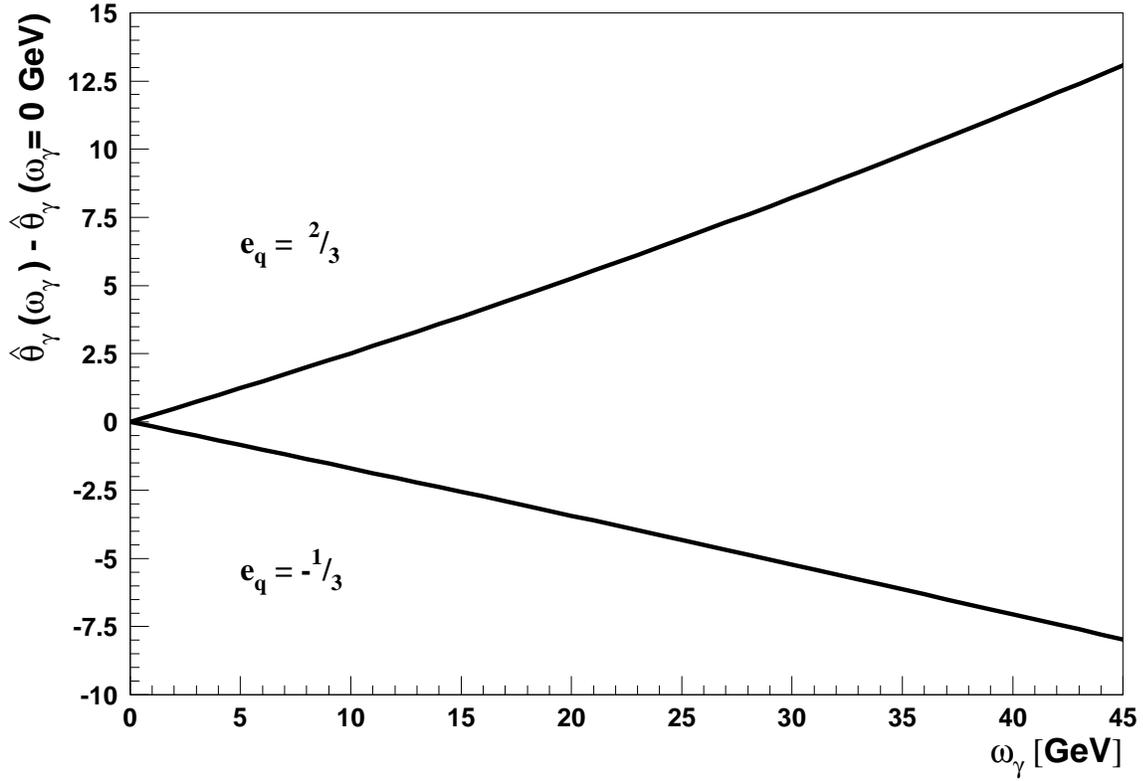,width=16.5cm}}
\caption[]{
The  positions  $\wh{\theta}_\gamma$  of the  radiation  zeros for the
processes  $e^-e^+  \rightarrow  q_d  \bar{q}_d  \gamma$  and  $e^-e^+
\rightarrow  q_u \bar{q}_u  \gamma$ as a function of the photon energy
$\omega_\gamma$,  for  $E_{e^-} = \sqrt{  {s}}/2 = 100$~GeV  and fixed
c.m.s.  frame (quark) scattering angle $\Theta_{\rm cm} = 90^\circ$.
}
\label{fig:nonsoft}
\end{center}
\end{figure}

To quantify this, we study planar  $e^-e^+ \to q \bar q \gamma$ events
in which  (i) the  polar  angle of the  quark  ($\Theta_{\rm  cm}$) is
fixed, (ii) the energy of the photon  ($\omega_\gamma$)  is fixed, and
(iii) the polar angle of the photon ($\theta_\gamma$) is varied.  Note
that the energy of the quark and the  four--momentum  of the antiquark
are  then  fixed  by  energy--momentum   conservation.  In  the  limit
$\omega_\gamma \to 0$ the kinematics of the soft--photon approximation
studied  in  previous   sections  are   reproduced.  We  find,  as  in
Ref.~\cite{Hey97b},  that the matrix  element has radiation  zeros for
non--zero  $\omega_\gamma$,  and that the  position of the zero varies
smoothly as $\omega_\gamma$  increases from zero.  This is illustrated
in     Fig.~\ref{fig:nonsoft},     which     shows    the     position
$\wh{\theta}_\gamma$ of the zero as a function of $\omega_\gamma$, for
$d$--type and $u$--type  quarks and $\Theta_{\rm cm} = 90^\circ$.  The
values  at   $\omega_\gamma   =  0$  coincide   with  those   obtained
analytically  in  the  soft--photon  approximation,  see  for  example
Fig.~\ref{fig:nomasspos}.  A  variation  of the  position  of the zero
with  the  photon  energy  is to be  expected,  since  with the  above
kinematics the direction of the antiquark changes as the photon energy
is varied.

If the photon is too energetic then the zeros can disappear.  This was
also  a  feature  of  the  $eq\to   eq\gamma$   process   studied   in
Ref.~\cite{Hey97b}.  For  example,  for $e_q = +2/3$ and  $\Theta_{\rm
cm} = 90^\circ$ we only have radiation zeros for $\omega_\gamma/E_{\rm
beam} < 0.47$.  However because of the soft--photon  energy  spectrum,
such upper  limits are not  particularly  relevant in practice.  Since
the position of the zero varies with the photon energy, any binning in
this    quantity    (above    say   some   small    threshold    value
$\omega_\gamma^{\rm  min}$) will remove the zero and replace it with a
sharp   minimum   located   near   the   corresponding    soft--photon
approximation  position.  We will  illustrate  this  in the  following
section.


\section{A Monte Carlo study for $b \bar b\gamma$ production}
\label{sec:montecarlo}

Our study so far has been based on the ideal but unrealistic situation
of well--defined  four--momenta  for the jets and the photon, fixed at
particular  directions in phase space.  In practice,  experiments deal
with  binned  quantities  and jets of finite  mass and  width.  A more
realistic study should therefore take these into account.  Rather than
try to model a particular detector  capability, we can define a simple
set of cuts which should take the main effects of smearing and binning
into  account.  The aim is to see whether the  radiation  zeros remain
visible after a more realistic  analysis.  We will,  however, make the
assumption  that in our sample of $b \bar b\gamma$ events the $b$--jet
can be  distinguished  from  the  $\bar  b$--jet.  This  guarantees  a
radiation  zero  in the  ideal  case,  as  discussed  in the  previous
sections.

We first  generate a sample of $b \bar b \gamma$  events using a Monte
Carlo which  includes  the exact phase  space and matrix  element.  We
choose a  centre--of--mass  energy of  $\sqrt{s} = 200\gev$.  For this
energy we can safely use the $m_b = 0$ massless  quark  approximation.
As a further  simplification  we include only $s$--channel  $\gamma^*$
exchange.\footnote{Including   also  $Z$  exchange  only  affects  the
overall normalisation and not the shape of the photon  distributions.}
The following sequence of cuts is applied:
\beq
10\gev < \omega_\gamma < 40\gev < E_{\bar b} < E_b \; ,
\eeq
to ensure that the photon is the softest  particle in the final state,
and that the  $b$--quark  direction  coincides with the thrust axis of
the event.  The photon is also  required to be separated in angle from
the beam and jet directions:
\beq
\theta_{\gamma,{\rm beam}} > 20^{\circ}\; , \qquad
\theta_{\gamma,b} \theta_{\gamma,\bar b} 
>10^{\circ}\; .
\eeq
These cuts serve to define a `measurable'  sample of $b \bar b \gamma$
events.

To investigate the radiation zero we must introduce a planarity cut on
the $b \bar b \gamma$  final state.  We do this by requiring  that the
normals  to the two  planes  defined  by (i)  the  beam  and  outgoing
$b$--quark   directions  and  (ii)  the  $\bar  b$--quark  and  photon
directions   are   approximately   parallel:  \begin{equation}   \vert
\vec{n}_{13}   \cdot   \vec{n}_{4k}   \vert  >  \cos   20^{\circ}\;  ,
\end{equation}   using   the   notation   for   momenta   defined   in
Eq.~(\ref{eq:qqphoton}).  We  can   then   study   the   polar   angle
($\theta_\gamma$) distribution of the photon for various values of the
polar  angle  ($\Theta_{\rm  cm}$)  of  the  thrust  axis  ($b$--quark
direction)  with  respect  to the  beam  direction.  In  practice,  we
consider  a bin  centred  on  $\theta_b  =  \Theta_{\rm  cm}$ of width
$10^{\circ}$, i.e.  we integrate over \begin{equation} \Theta_{\rm cm}
-  5^{\circ}  <  \theta_b  <   \Theta_{\rm   cm}  +  5^{\circ}   \;  .
\end{equation} Note that our cuts are deliberately chosen to mimic the
soft--photon   kinematics  used  in  Section~2.  However   because  we
integrate  over the  photon  energy  and  smear  the  polar  angle and
planarity  criteria  we  expect  to  see  {\it  dips}  in  the  photon
distribution rather than strict zeros.

\begin{figure}[t]
\begin{center}
\mbox{\epsfig{figure=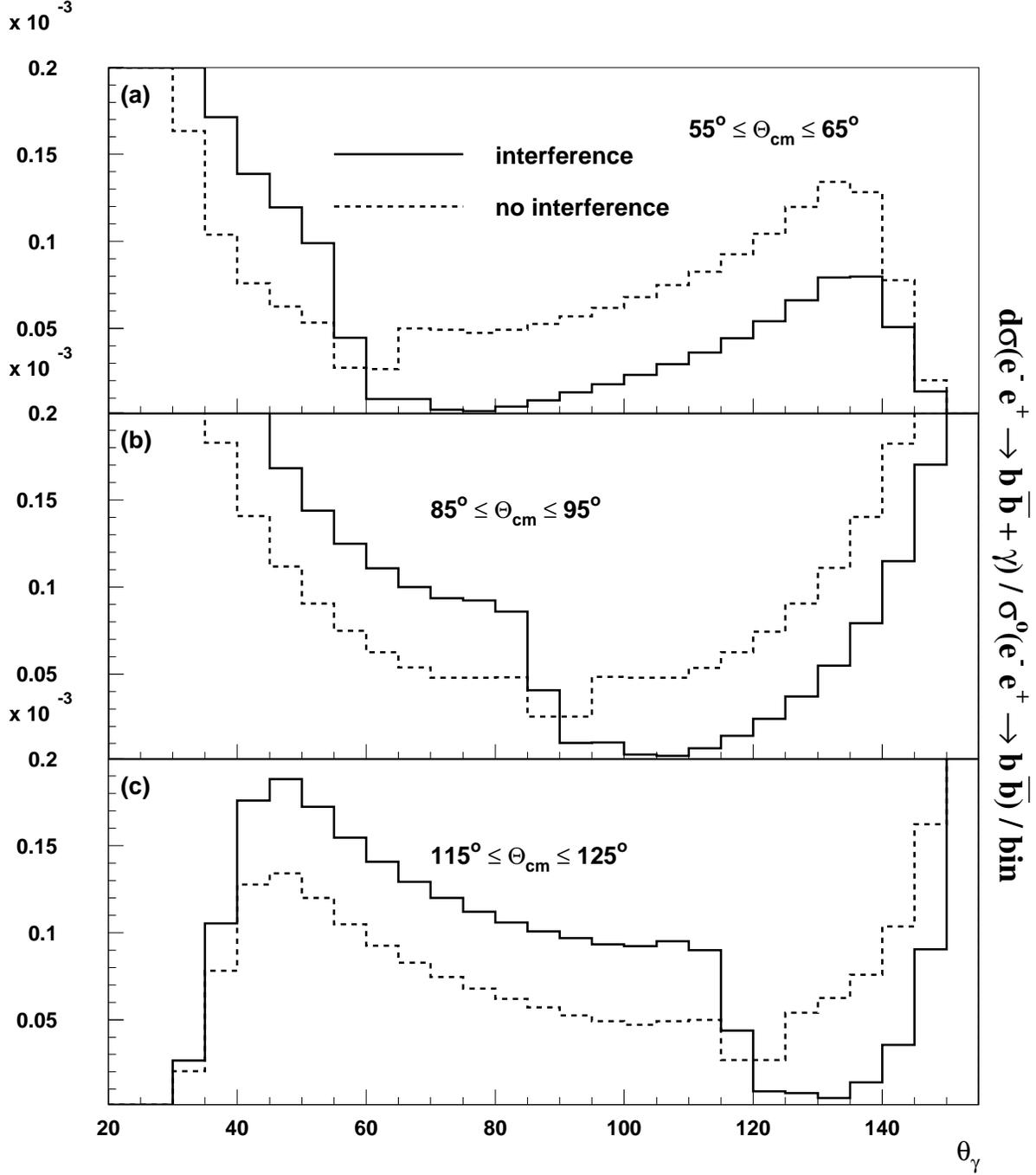,width=16.5cm}}

\caption[]{
The  $\theta_\gamma$  distribution (solid histograms)  obtained in the
Monte Carlo  calculation of $e^-e^+ \to b \bar b \gamma$ in the planar
configuration.  The  various   cuts  are  defined  in  the  text.  The
$\Theta_{\rm cm}$ angles are
                             (a) $60^\circ $,      
                             (b) $90^\circ $ 
                         and (c) $120^\circ$. 
The dashed lines are the results of the corresponding calculation with
the interference terms removed.
}
\label{fig:histo}
\end{center}
\end{figure}

Figure~\ref{fig:histo} shows the $\theta_\gamma$ distribution for 
(a) $\Theta_{\rm cm}  = 60^{\circ} $,  
(b) $\Theta_{\rm cm}  = 90^{\circ} $ and
(c) $\Theta_{\rm cm}  = 120^{\circ}$.
Comparing with Fig.~\ref{fig:nomassd}, we once again see sharp dips at
approximately the same position as in the `ideal'  soft--photon  case.
Note    that    the     collinear     singularities     evident     in
Fig.~\ref{fig:nomassd}  are now removed by the cuts.  The  suppression
of the cross  section  at the  position  of the zeros can  further  be
appreciated   by  comparing   with  the  results   obtained  when  the
interference  term  in the  matrix  element  squared  is set to  zero,
corresponding to incoherent  photon emission off the initial and final
states.  The results of this approximate  calculation, shown as dashed
lines in Fig.~\ref{fig:histo}, do not exhibit any dip structure in the
region of the zeros and are  clearly  distinguishable  from the  exact
results.

\section{Conclusions}
\label{sec:conc}

Radiation zeros are an important consequence of the gauge structure of
the  electromagnetic  interaction.  They arise in  different  types of
high--energy scattering processes.  In this paper we have investigated
a particular type of radiation zero (`type 2' or `planar')  which is a
feature  of the  process  $e^+e^-  \to q \bar q  \gamma$.  We  derived
expressions for the locations of the zeros in the soft--photon  limit,
and showed that the zeros persist for hard photons and massive quarks.
However   the   experimental   verification   of  such  zeros  is  not
straightforward.  The zeros  disappear  on the $Z^0$ pole  because the
interference   between   initial--  and   final--state   radiation  is
suppressed  by the finite $Z$  lifetime.  The  collision  energy  must
therefore be greater or less than $M_Z$.  Unfortunately  the number of
events beyond the $Z^0$ pole at present colliders is quite low.  Apart
from the resulting issue of the overall event rate, it is necessary to
be able to  distinguish  quark from antiquark jets in order to compare
with our  predictions.  This can perhaps we done with some  efficiency
for  $b$--quark  jets.  We performed a Monte Carlo study which  showed
that  `realistic'  distributions  do  indeed  exhibit  sharp  dips  in
particular  regions  of  phase  space.  Further  studies  using a more
complete simulation of the final--state hadronisation process would be
worthwhile.

\vspace{1.0cm}

\noindent {\bf Acknowledgements} \\

We are grateful to Z.~W\c{a}s for  discussions on $Z^0$ pole radiation
and  J.~Turnau  for some  helpful  remarks.  MH  wishes  to thank  the
members  of  the  {\em  H.~Niewodnicza{\'n}ski  Institute  of  Nuclear
Physics} (Krak\'ow) for their  hospitality  during the final stages of
this work and also gratefully  acknowledges  financial  support in the
form of a DAAD--Dok\-tor\-an\-den\-sti\-pen\-di\-um (HSP III).

\newpage



\newpage


\end{document}